\documentclass[aps,prl,twocolumn,showpacs,citeautoscript,reprint,longbibliography,superscriptaddress]{revtex4-1}
\usepackage{boldline,multirow}
\usepackage{float}
\usepackage{graphicx,natbib}
\usepackage{bm,amsmath,amssymb,wasysym,amsfonts,dcolumn}
\usepackage[colorlinks=true, urlcolor=blue, linkcolor=blue, citecolor=blue, pdfborder={0 0 0}]{hyperref}
\usepackage[usenames,dvipsnames]{xcolor}
\usepackage{lipsum}
\usepackage{epstopdf}
\usepackage{cleveref}
\usepackage[T1]{fontenc}  
\crefname{figure}{Fig.}{Figs}
\crefname{table}{Table}{Tables}
\crefname{section}{Sec.}{Section}

\begin{document}

\title{Spin-flip diffusion length in 5d transition metal elements: a first-principles benchmark}

\author{Rohit S.Nair}
\affiliation{Faculty of Science and Technology and MESA$^+$ Institute for Nanotechnology, University of Twente, P.O. Box 217,
		7500 AE Enschede, The Netherlands}
		
\author{Ehsan Barati}
\affiliation{Faculty of Science and Technology and MESA$^+$ Institute for Nanotechnology, University of Twente, P.O. Box 217,
		7500 AE Enschede, The Netherlands}		
\altaffiliation[Present address:]{Department of Chemistry, Brown University, Providence, RI 02912, United States}

\author{Kriti Gupta}
\affiliation{Faculty of Science and Technology and MESA$^+$ Institute for Nanotechnology, University of Twente, P.O. Box 217,
		7500 AE Enschede, The Netherlands}

\author{Zhe Yuan}
\email[Email: ]{zyuan@bnu.edu.cn}
\affiliation{The Center for Advanced Quantum Studies and Department of Physics, Beijing Normal University, 100875 Beijing, China}
		
\author{Paul J. Kelly}
\email[Email: ]{P.J.Kelly@utwente.nl}
\affiliation{Faculty of Science and Technology and MESA$^+$ Institute for Nanotechnology, University of Twente, P.O. Box 217,
		7500 AE Enschede, The Netherlands}
\affiliation{The Center for Advanced Quantum Studies and Department of Physics, Beijing Normal University, 100875 Beijing, China}		

\date{\today}

\begin{abstract}
Little is known about the spin-flip diffusion length $l_{\rm sf}$, one of the most important material parameters in the field of spintronics. 
We use a density-functional-theory based scattering approach to determine values of $l_{\rm sf}$ that result from electron-phonon scattering as a function of temperature for all 5$d$ transition metal elements.
$l_{\rm sf}$ does not decrease monotonically with the atomic number $Z$ but is found to be inversely proportional to the density of states at the Fermi level. By using the same local current methodology to calculate the spin Hall angle $\Theta_{\rm sH}$ that characterizes the efficiency of the spin Hall effect, we show that the products $\rho(T) \, l_{\rm sf}(T)$ and $\Theta_{\rm sH}(T) \, l_{\rm sf}(T)$ are constant.  
\end{abstract}


\maketitle

Spin-orbit coupling (SOC) leads to the loss of spin angular momentum. A current of electrons injected from a ferromagnet into a nonmagnetic material loses its spin-polarization over a length scale of $l_{\rm sf}$, the spin-flip diffusion length (SDL) \cite{Aronov:jetpl76, Johnson:prl85, vanSon:prl87, *vanSon:prl88, Valet:prb93}, making the observation of spin currents difficult. The giant magnetoresistance (GMR) effect was discovered in magnetic multilayers \cite{Baibich:prl88, Binasch:prb89} only when the thickness of the spacer layers separating the magnetic films was made to be of order $l_{\rm sf}$. A review of the SDL in metals and alloys some twenty years after the discovery of GMR concerned mainly the free-electron like metals Cu, Ag, Au and Al that have large values of $l_{\rm sf}$;  for just a few of the transition metal elements, there was a single, low temperature entry \cite{Bass:jpcm07}. With the ``rediscovery'' \cite{Hirsch:prl99, Zhang:prl00} of the spin Hall effect (SHE) \cite{Dyakonov:zetf71, *Dyakonov:pla71} and its observation in semiconductors \cite{Kato:sc04, Wunderlich:prl05} and metals \cite{Valenzuela:nat06, Saitoh:apl06}, this situation  has changed radically \cite{Hoffmann:ieeem13, Sinova:rmp15}. However, even for well-studied materials like Pt, values of $l_{\rm sf}$ reported over the last decade span an order of magnitude \cite{Sinova:rmp15, Wesselink:prb19}. Discernible trends in $l_{\rm sf}$ have not been reported for different transition metal elements. 
 
The SHE is another consequence of SOC whereby the passage of a charge current through a metal gives rise to a transverse spin current that can enter an adjacent magnetic material and exert a torque on its magnetization causing it to switch its orientation. The efficiency of the SHE is given by the spin Hall angle (SHA) $\Theta_{\rm sH}$ that is the ratio of the transverse spin current (measured in units of $\hbar/2$) to the charge current (measured in units of the electron charge $-e$). From being a curiosity, the SHE has rapidly become a leading contender to form the basis for a new magnetic memory technology \cite{Bhatti:mt17} bringing with it the need to find materials with optimal values of $\Theta_{\rm sH}$ with a primary focus on heavy metals like Pt \cite{Saitoh:apl06, Kimura:prl07}, Ta \cite{Liu:sc12} and W \cite{Pai:apl12}. Striking discrepancies between different room temperature (RT) measurements, with reported values of e.g. $\Theta_{\rm sH}^{\rm Pt}$ ranging between 1\% and 11\% \cite{Hoffmann:ieeem13}, led to the realization that the bulk parameters $l_{\rm sf}$ and $\Theta_{\rm sH}$ as well as the resistivity $\rho$ were very sample dependent and needed to be determined simultaneously. Doing this did not however lead to a consensus about the values of these parameters \cite{Sinova:rmp15}. Whether the SHA is determined using spin pumping and the inverse SHE \cite{Saitoh:apl06, Ando:prl08, Mosendz:prl10, Mosendz:prb10}, the SHE and spin-transfer torque \cite{Liu:prl11}, or nonlocal spin-injection \cite{Kimura:prl07, Vila:prl07}, interfaces are always involved leading to the suggestion that interface processes like interface spin flipping (spin memory loss) \cite{Rojas-Sanchez:prl14, LiuY:prl14, Nguyen:prl16, Sagasta:prb16} or an interface SHE \cite{WangL:prl16, Amin:prb16a} should be taken into consideration in interpreting experiment. Attempts to do so have, if anything, made matters worse with recently determined values of $l_{\rm sf}^{\rm Pt}$ ranging from 1.4 to 11~nm and $\Theta_{\rm sH}^{\rm Pt}$ ranging between 3\% and 39\%; see Table~V in Ref.~\cite{Wesselink:prb19}. 

The only attempt we are aware of to study the intrinsic SDL theoretically is the {\color{blue}phonon-induced spin relaxation work} by Fabian and Das Sarma on aluminium \cite{Fabian:prl99} that is not readily generalized to transition metals. The purpose of this Letter is to present benchmark calculations of $l_{\rm sf}$ and $\Theta_{\rm sH}$ for all bulk 5$d$ metals at temperatures where the resistivity is dominated by electron-phonon scattering and the extrinsic scattering that dominates low temperature measurements, but whose microscopic origin is seldom known, can be disregarded. These first quantitative theoretical predictions give us insight into how $l_{\rm sf}$ may be modified and confirm long-standing speculations about relationships between $\rho$, $l_{\rm sf}$, $\Theta_{\rm sH}$ and  the temperature $T$.

\begin{figure}[t]
\includegraphics[width=8.4 cm]{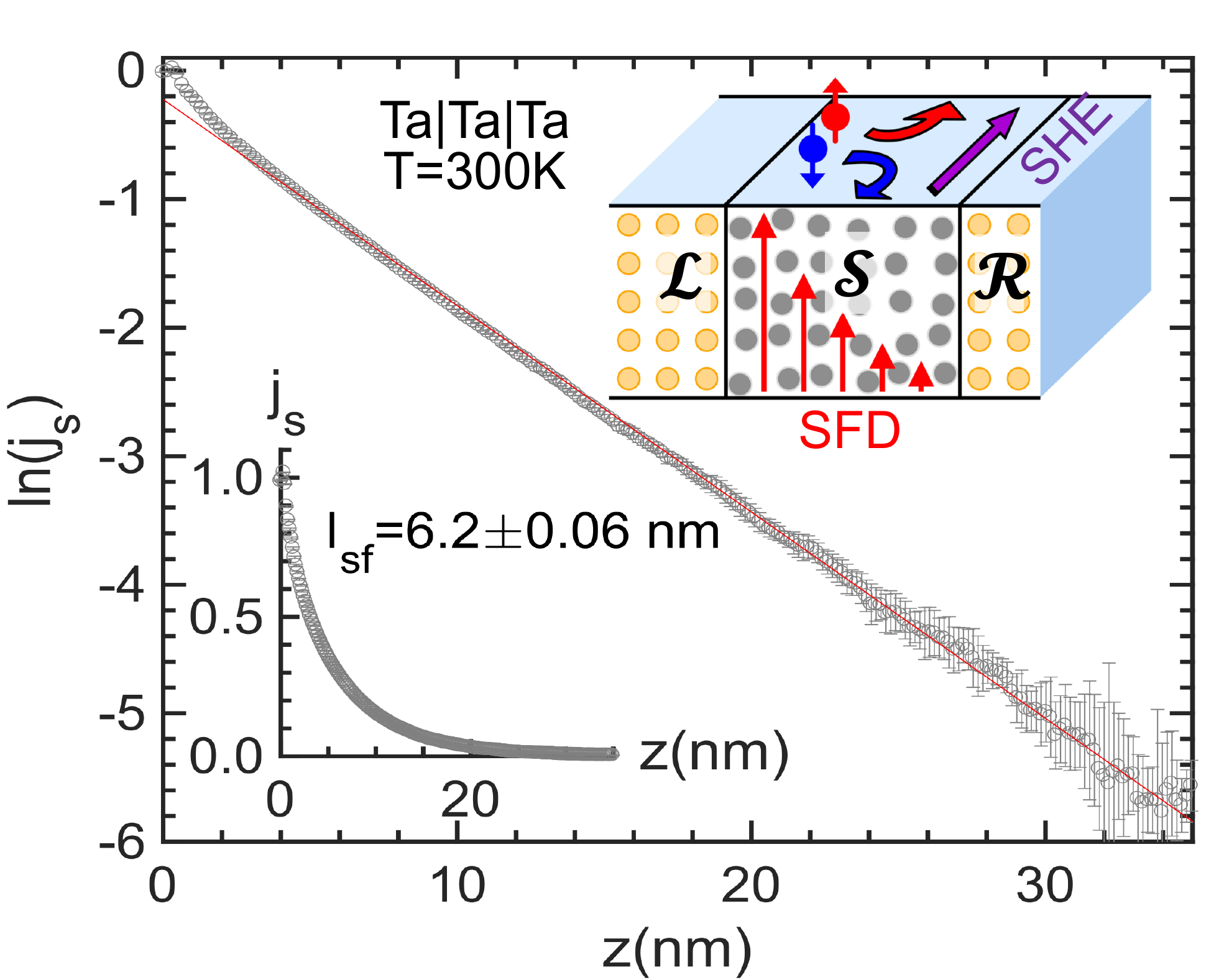} 
\caption{Natural logarithm of a spin current injected into RT bcc Ta as a function of the coordinate $z$ in the transport direction, $\mathcal{L}$$\rightarrow$$ \mathcal{R}$. 
The upper inset sketches the transport geometry with a scattering region $\mathcal{S}$ with temperature dependent lattice disorder sandwiched between ideal semi-infinite ballistic leads, $\mathcal{L}$ and $\mathcal{R}$. A fully polarized spin current injected from the left lead $\mathcal{L}$ undergoes spin flipping leading to spin equilibration on a length scale given by the spin-flip diffusion (SFD) length $l_{\rm sf}$ as suggested by the red arrows. An unpolarized charge current injected from $\mathcal{L}$ undergoes spin dependent scattering leading to a transverse spin Hall current depicted by the purple arrow.
The lower inset shows the spin current on a linear scale. 
The current was extracted from the results of a scattering calculation for a two-terminal Ta$\uparrow$$|$Ta$|$Ta configuration using a $7 \times 7$ lateral supercell where Ta$\uparrow$ indicates an artificial, fully polarized Ta lead. The red line is a weighted linear least squares fit; the error bar in the value $6.20\pm0.06$ results from different ``reasonable'' weightings and cutoff values \cite{Wesselink:prb19}. 
}
\label{Fig1}
\end{figure}

{\color{red}\it Method.---}We recently demonstrated that it was possible to distinguish bulk transport properties from the interface effects that are inherent to scattering formulations of transport \cite{Datta:95} by evaluating local charge and spin currents from the solutions of quantum mechanical scattering calculations \cite{Wesselink:prb19, Gupta:prl20} as sketched in the inset to \cref{Fig1}. Our density-functional-theory scattering calculations include temperature-induced lattice and spin disorder {\color{blue}in the adiabatic approximation} \cite{LiuY:prb11, LiuY:prb15} as well as SOC \cite{Starikov:prb18}. An example of the results of such a calculation is shown in \cref{Fig1} where we plot the natural logarithm of the spin current that results from injecting a fully polarized current into a length $L$ of room-temperature (RT, $T=300 \,$K) disordered Ta {\color{blue}where a Gaussian distribution of random atomic displacements was chosen to reproduce the experimentally observed resistivity of bulk Ta \cite{HCP90}.} From the near-perfect exponential decay over five orders of magnitude, we extract a value of $l_{\rm sf}^{\rm Ta}(T=300 \,{\rm K}) \sim 6.2\,$nm {\color{blue}that is independent of the lead material \cite{Wesselink:prb19}}. Spin memory loss at the left interface manifests itself in the deviation from exponential behaviour close to $z \sim 0$ \cite{Gupta:prl20}.

\begin{figure}[t]
\includegraphics[width=8.6cm]{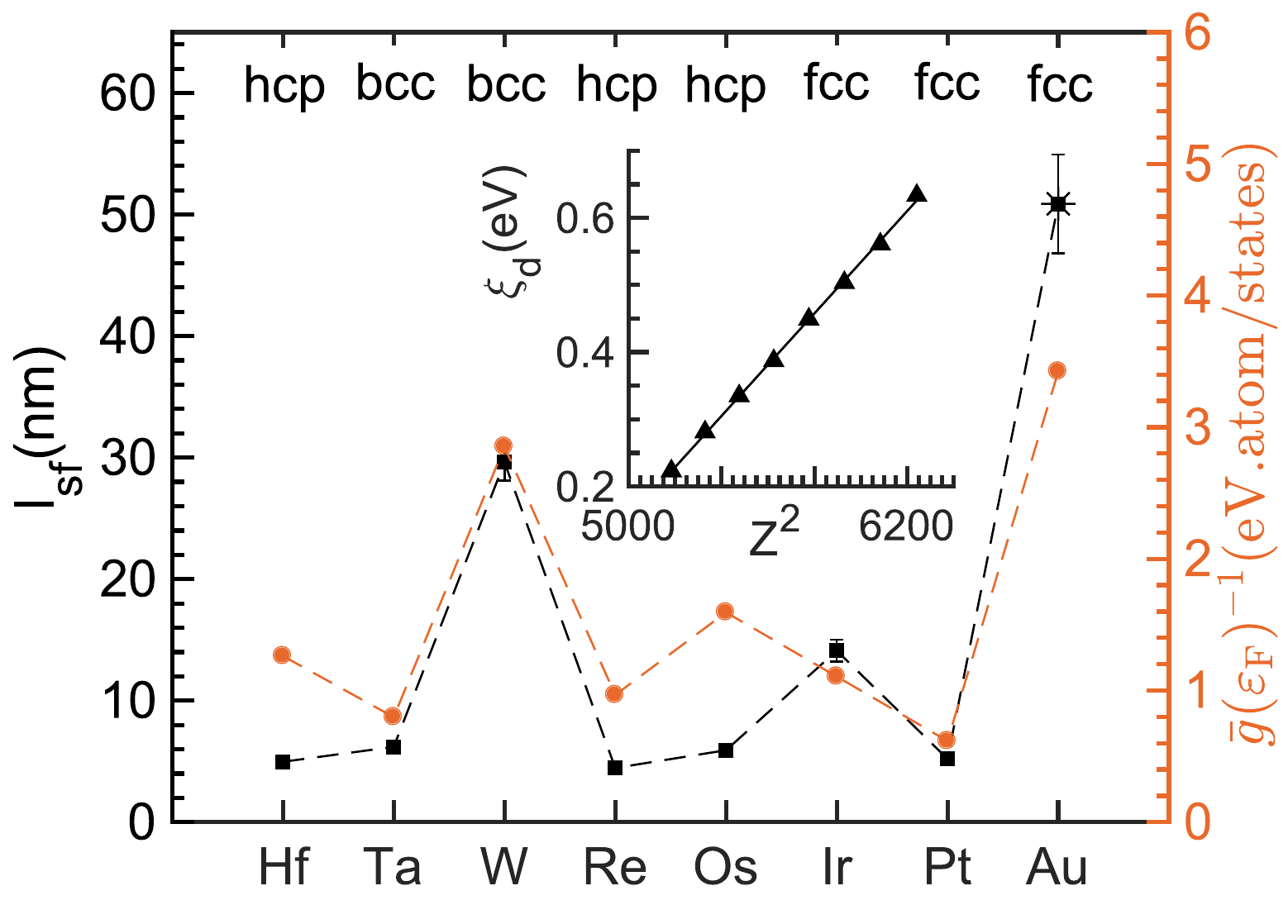}  
\caption{Black (lhs): spin-flip diffusion length $l_{\rm sf}$ for 5$d$ transition metals calculated at room temperature (300K), the error bars correspond to the spread of values for ten different configurations. For hcp metals, the $c$-axis values are shown. Orange (rhs): inverse of $\bar{g}(\varepsilon_F)$, the density of states averaged over an energy window of $\pm k_B T$ about the Fermi energy $\varepsilon_F$. Inset: spin-orbit coupling parameter $\xi_d$ as a function of the square of the atomic number $Z$ for the $5d$ elements. 
}
\label{Fig2}
\end{figure}

{\color{red}\it Results.---}The parameter $\xi$ contained in the SOC term $\xi {\bf l.s}$ of the Pauli Hamiltonian scales as the square of the atomic number $Z$ \cite{Herman:63, Mackintosh:80} as shown in the inset to \cref{Fig2}. For the $5d$ electrons, $\xi_d$ increases monotonically from $\sim 0.22 \,$eV for Hf to $\sim 0.63 \,$eV for Au. This might lead us to expect a decreasing trend in $l_{\rm sf}$ with $Z$. The calculated RT values of $l_{\rm sf}$ shown in \cref{Fig2} for all 5$d$ elemental metals exhibit no such decrease. For example, Ta and W are both bcc and as neighbours in the periodic table have very similar values of $\xi_d$ yet $l_{\rm sf}^{\rm W}$ is some five times larger than $l_{\rm sf}^{\rm Ta}$. We find instead that the dominant trend is given by the inverse of the Fermi level density of states (DoS),  $g(\varepsilon_F)$, shown in orange in \cref{Fig2} averaged over an energy window of $\pm k_BT$ about the Fermi energy. A low DoS means fewer possibilities to scatter with a spin flip. The same correlation with the inverse DoS is found on calculating $l_{\rm sf}$ as a function of band-filling for bcc and fcc structures (not shown). Reported experimental values of $l_{\rm sf}$ at RT are 1.8 and 1.9 nm for bcc Ta, 2.1 nm for bcc W, between 1 and 11 nm for Pt and from 27 to 86 nm for Au \cite{Sinova:rmp15}. The lack of any correlation with the ``intrinsic'' values we calculate suggests that the measurements are dominated by other, as yet unidentified, factors.   

\begin{figure}[t]
\includegraphics[width=8.4 cm]{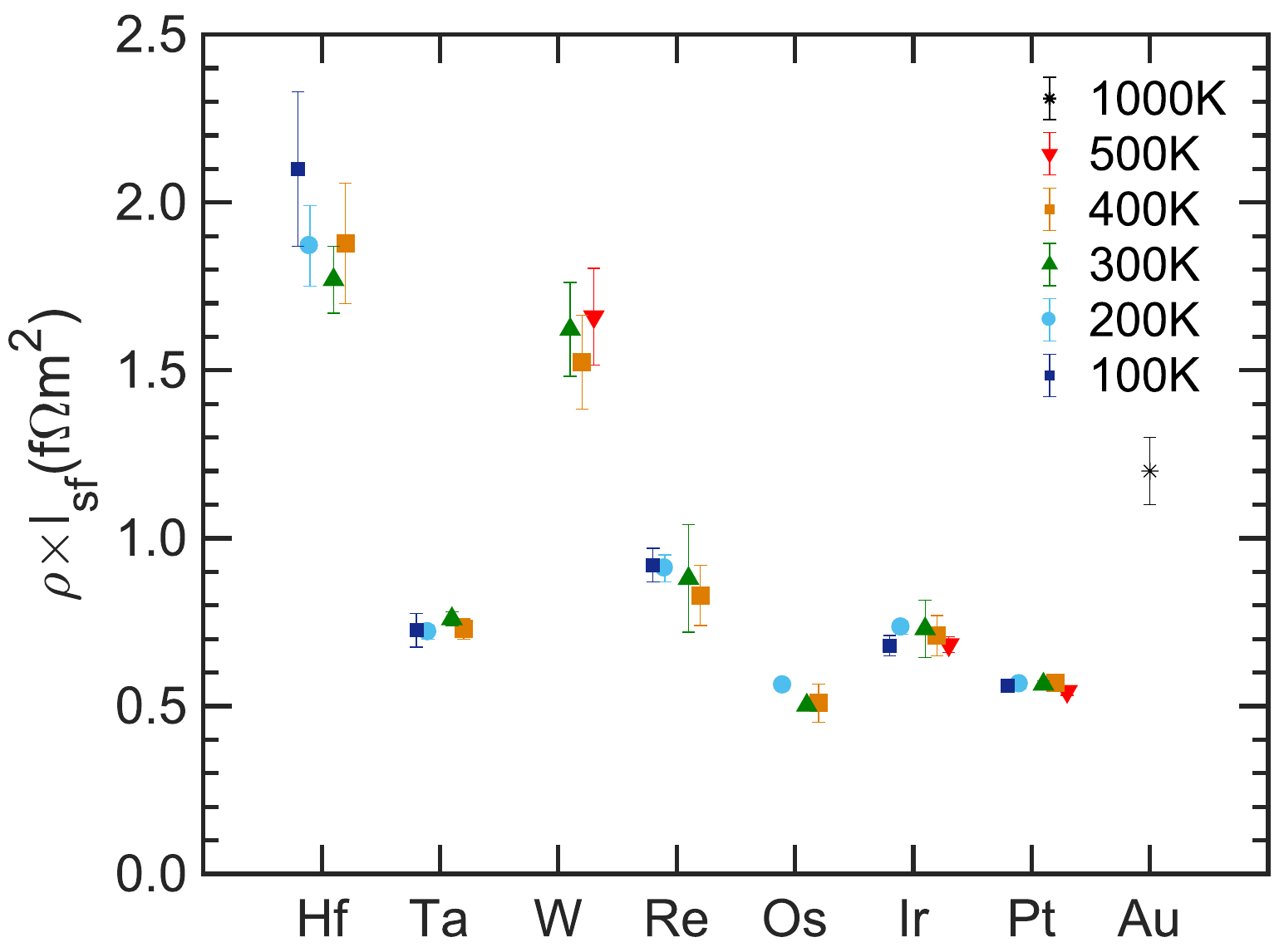} 
\caption{Product of the spin-flip diffusion length and resistivity for 5$d$ metals calculated at different temperatures. For hcp metals, the $c$-axis values are shown.  The error bars are estimated from the average spreads over 10 configurations for both $l_{\rm sf}$ and $\rho$.
 }
\label{Fig3}
\end{figure}

{\color{red}\it Temperature dependence.---}By varying the mean square displacement of the atoms in the scattering region to reproduce experimental resistivities, we can study the temperature dependence of $l_{\rm sf}$ due to electron-phonon coupling. For temperatures in the range 100-500~K, the product $\rho(T) \, l_{\rm sf}(T)$ is plotted for all 5$d$ elements in \cref{Fig3} where it is seen to be independent of temperature within the error bars of the calculations. 
(The large value of $l_{\rm sf}^{\rm W} \sim 30 \,$nm at 300K requires calculations with an excessively long geometry putting values for 100 and 200 K out of reach.)
This is in agreement with predictions made by Elliott and Yafet for doped semiconductors and alkali elements \cite{Elliott:pr54, Yafet:63} but now for Fermi surfaces that are far more complex than those they considered, for which their approximations are not applicable. As such, this result is nontrivial. 
While $l_{\rm sf}$ varies from 4 to 50~nm at room temperature, the product $\rho l_{\rm sf}$ spans a smaller range, varying between 0.5 and 2 f$\Omega$m$^2$.

Although $\xi_d$ attains its maximum value for Au, the $d$ bands are then completely filled and well below the Fermi energy so Au is expected to have a long SDL. The low resistivity and weak effective SOC of Au make a direct calculation computationally very challenging. To determine  $l_{\rm sf}^{\rm Au}$ at room temperature, we considered an elevated temperature of $T=1000 \,$K and then assumed that $\rho(T) \, l_{\rm sf}(T)$ was a constant in order to estimate the RT value of $l_{\rm sf} \sim 50 \,$nm shown in \cref{Fig2} and given in \cref{tab:parameters}.

\begin{figure}[b]
\includegraphics[width=8.4 cm]{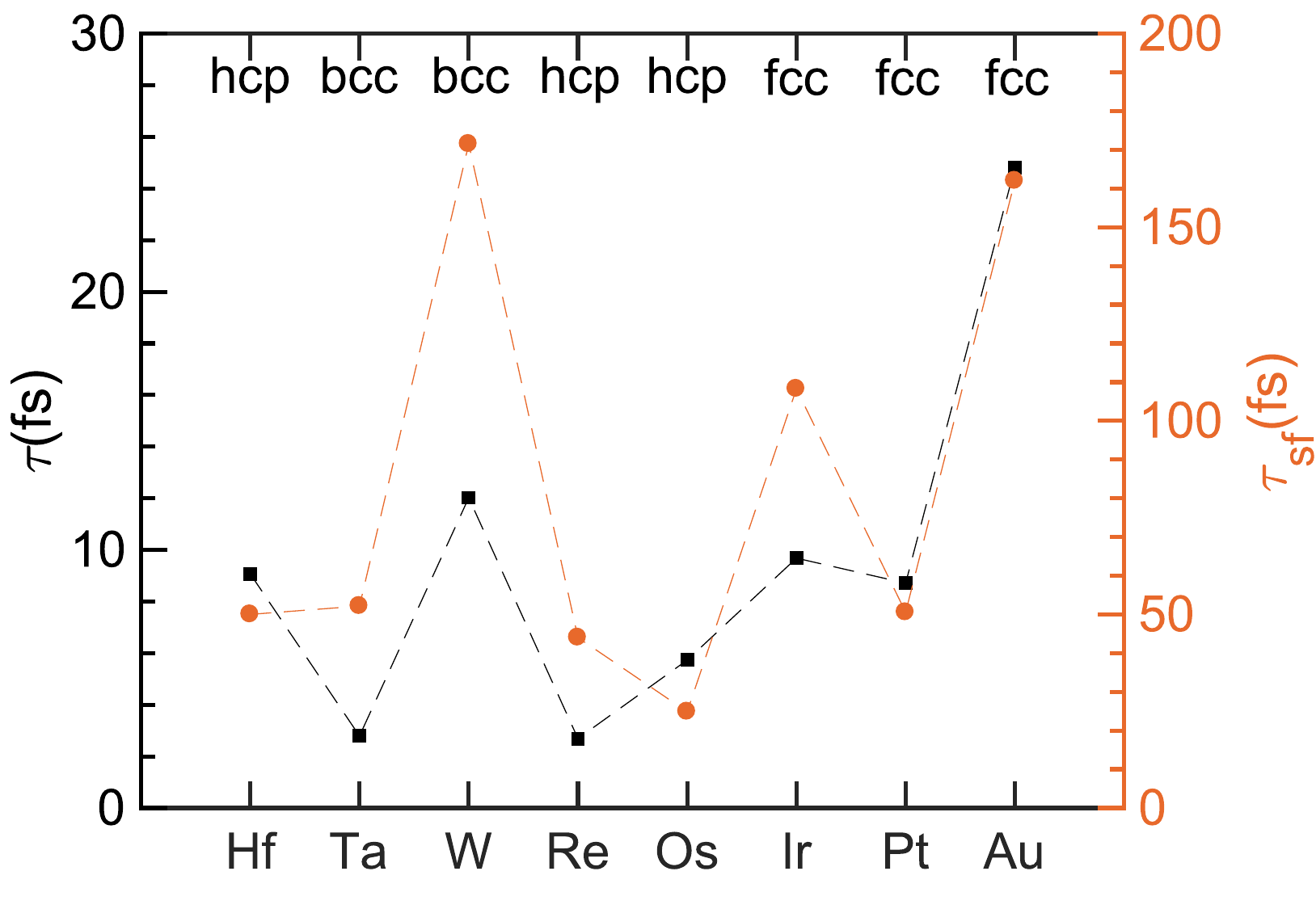} 
\caption{Momentum relaxation time $\tau$ and spin-flip relaxation time $\tau_{\rm sf}$ in fs for all 5$d$ elements at room temperature. 
}
\label{Fig4}
\end{figure}

{\color{red}\it Relationship of $\tau_{\rm sf}$ to $\tau$.---}The product $\rho(T) \, l_{\rm sf}(T) $ is expected to be a constant when momentum scattering is dominated by phonons and $\tau_{\rm sf} \,$$\propto$$\, \tau$ \cite{Elliott:pr54, Yafet:63}. We can estimate $\tau$ and $\tau_{\rm sf}$ as follows.
In the relaxation time approximation, the conductivity is given in terms of the ${\bf k}$ dependent velocities ${\bm \upsilon_n({\bf k})}=\frac{1}{\hbar} \nabla_{\bf k}\varepsilon_n({\bf k})$ for band $n$ as
\begin{equation}
\sigma_{ij}= e^2 \sum_n \iiint \frac{d^3k}{8\pi^3} \tau_n({\bf k}) \,
           \upsilon_{ni}({\bf k})\upsilon_{nj}({\bf k}) 
           \Big(-\frac{\partial f}{\partial \varepsilon}\Big)_{\varepsilon=\varepsilon_n({\bf k})} \nonumber
\end{equation}
which becomes an integral over the Fermi surface $S_F$
when $-\frac{\partial f}{\partial \varepsilon} \rightarrow \delta(\varepsilon - \varepsilon_F)$ in the low temperature limit and assuming $\tau({\bf k})=\tau(\varepsilon_n({\bf k})) $ so
$\sigma =  e^2 g(\varepsilon_F) \tau(\varepsilon_F) \langle v^2_F \rangle . $
Both $g(\varepsilon_F)$ and $\langle v^2_F \rangle$ can be evaluated from standard bulk LMTO electronic structure calculations \cite{footnote3}.
Since $\sigma \equiv 1/\rho$ is known \cite{HCP90, footnote1}, $\tau$ can be evaluated as         
\begin{equation}
\tau = \frac{\sigma}{e^2 g(\varepsilon_F) \langle v^2_F \rangle} \; . \nonumber
\end{equation}
The diffusion coefficient $D$ \cite{vanSon:prl87} can be determined from the Einstein relation $\sigma = D e^2 g(\varepsilon_F)$. Using the spin-flip diffusion length $l_{\rm sf}$ evaluated from the exponential decay of an injected spin current and the relationship $ l_{\rm sf}^2=D\tau_{\rm sf}$ allows us to determine $\tau_{\rm sf}$. The ratio of the spin relaxation time, $\tau_{\rm sf}$, to $\tau$ is finally
\begin{equation}
\frac{\tau_{\rm sf}}{\tau}= \big[e^2 \rho \, l_{\rm sf}  g(\varepsilon_F) \big]^2 \langle v^2_F \rangle . \nonumber
\label{eqtauratio}
\end{equation}
In spite of the apparent complexity of this relationship, \cref{Fig4} shows that the factor dominating the $Z$ dependence of both relaxation times is the inverse Fermi-level density of states. In this sense $\tau_{\rm sf} \,$$\propto$$\, \tau$, consistent with $\rho \, l_{\rm sf} $ being independent of temperature. We note that the electron-phonon coupling and phonon-modulated SOC effects enter our calculation implicitly via the resistivity $\rho$ while the density of states $g(\varepsilon_F)$ is calculated with SOC explicitly included in the Hamiltonian. Room temperature values of all the transport parameters calculated as described above are given in \cref{tab:parameters}.


\begin{table}[t]
\caption{Room temperature transport parameters for 5$d$ metals: 
resistivity $\rho$ ($\mu\Omega \,$cm); 
Fermi level density of states $\bar{g}(\varepsilon_F)$ (states/eV.atom);
diffusion constant $D$ (cm$^2/$s);
Fermi velocity $\upsilon_F \equiv \langle \upsilon_F^2 \rangle^\frac{1}{2}$ ($10^8{\rm cm/s}$); 
relaxation time $\tau$ (fs);
spin-flip diffusion length $l_{\rm sf}$ (nm);
spin relaxation time $\tau_{\rm sf}$ (fs);  
spin Hall angle $\Theta_{\rm sH}$ (\%). 
$\parallel$ and $\perp$ refer to parallel to and perpendicular to the hcp hexagonal axis, respectively.}
\begin{ruledtabular}
\begin{tabular}{llclccccrr}
El. & Latt. & \multicolumn{1}{c}{$\rho$} 
            & \multicolumn{1}{c}{$\bar{g}(\varepsilon_F)$} 
            & \multicolumn{1}{c}{$D$}
            & \multicolumn{1}{c}{$\upsilon_F $} 
            & \multicolumn{1}{c}{$\tau$} 
            & \multicolumn{1}{c}{$l_{\rm sf}$} 
            & \multicolumn{1}{c}{$\tau_{\rm sf}$} 
            & \multicolumn{1}{c}{$\Theta_{\rm sH}$} \\ 
\cline{1-10}	
Hf & hcp $\parallel$ & 35.6  & 0.79 &   4.94 & 0.40 &  9.07 &  4.97 &  50 &  1.35 \\
   & hcp $\perp$     & 59.0  &      &   2.98 &      &  5.47 &  4.2  &  60 &  0.98 \\
Ta & bcc             & 12.1  & 1.25 &   7.37 & 0.89 &  2.79 &  6.16 &  50 & -0.50 \\
 W & bcc             &  5.49 & 0.35 &  51.2  & 1.13 & 12.0  & 29.6  & 170 & -0.40 \\
Re & hcp $\parallel$ & 19.7  & 1.04 &   4.52 & 0.71 &  2.69 &  4.46 &  44 & -1.28 \\
   & hcp $\perp$     & 25.8  &      &   3.45 &      &  2.05 &  3.22 &  30 & -1.90 \\
Os & hcp $\parallel$ &  9.48 & 0.63 &  14.77 & 0.88 &  5.73 &  6.06 &  25 &  0.71 \\
   & hcp $\perp$     & 10.0  &      &   3.41 &      &  5.44 &  5.90 &  25 & -0.66 \\
Ir & fcc             &  5.31 & 0.90 &  18.4  & 0.76 &  9.68 & 14.1  & 110 &  0.22 \\
Pt & fcc             & 10.8  & 1.62 &   5.37 & 0.43 &  8.71 &  5.21 &  50 &  4.02 \\
Au & fcc             &  2.27 & 0.29 & 160    & 1.39 & 24.8  & 50.9  & 160 &  0.25 \\
\end{tabular}
\end{ruledtabular}
\label{tab:parameters}
\end{table}

{\color{red}\it Spin Hall Angle.---}Because of the correlation between the SHA and SDL observed in measurements \cite{Rojas-Sanchez:prl14}, it is desirable to determine $\Theta_{\rm sH}$ using the same approximations as were used to calculate $l_{\rm sf}$.  {\color{blue}Most quantitative theoretical studies of the SHE \cite{Guo:prl08, Tanaka:prb08, Qiao:prb18, Ryoo:prb19} are based upon the Kubo formalism and have focused on the so-called ``intrinsic'' contribution that does not consider the role of the electron-phonon scattering mechanism that dominates the resistivity of elemental metals at room temperature where the vast majority of $\Theta_{\rm sH}$ determinations have been made \cite{Sinova:rmp15}. In the linear response regime, the scattering theory we use is equivalent to the Kubo formalism \cite{Khomyakov:prb05} and therefore includes the ``intrinsic'' contribution as well as that from electron-phonon coupling. The advantage of the scattering formalism is that extrinsic mechanisms can be included on an equal footing.} By calculating the transverse spin current resulting from a longitudinal charge current using the local current method \cite{Wesselink:prb19} introduced previously to study $\Theta_{\rm sH}^{\rm Pt}$ \cite{WangL:prl16}, {\color{blue}we determined $\Theta_{\rm sH}$ as a function of temperature for all 5$d$ elements.}

\begin{figure}[t]
\includegraphics[width=8.4 cm]{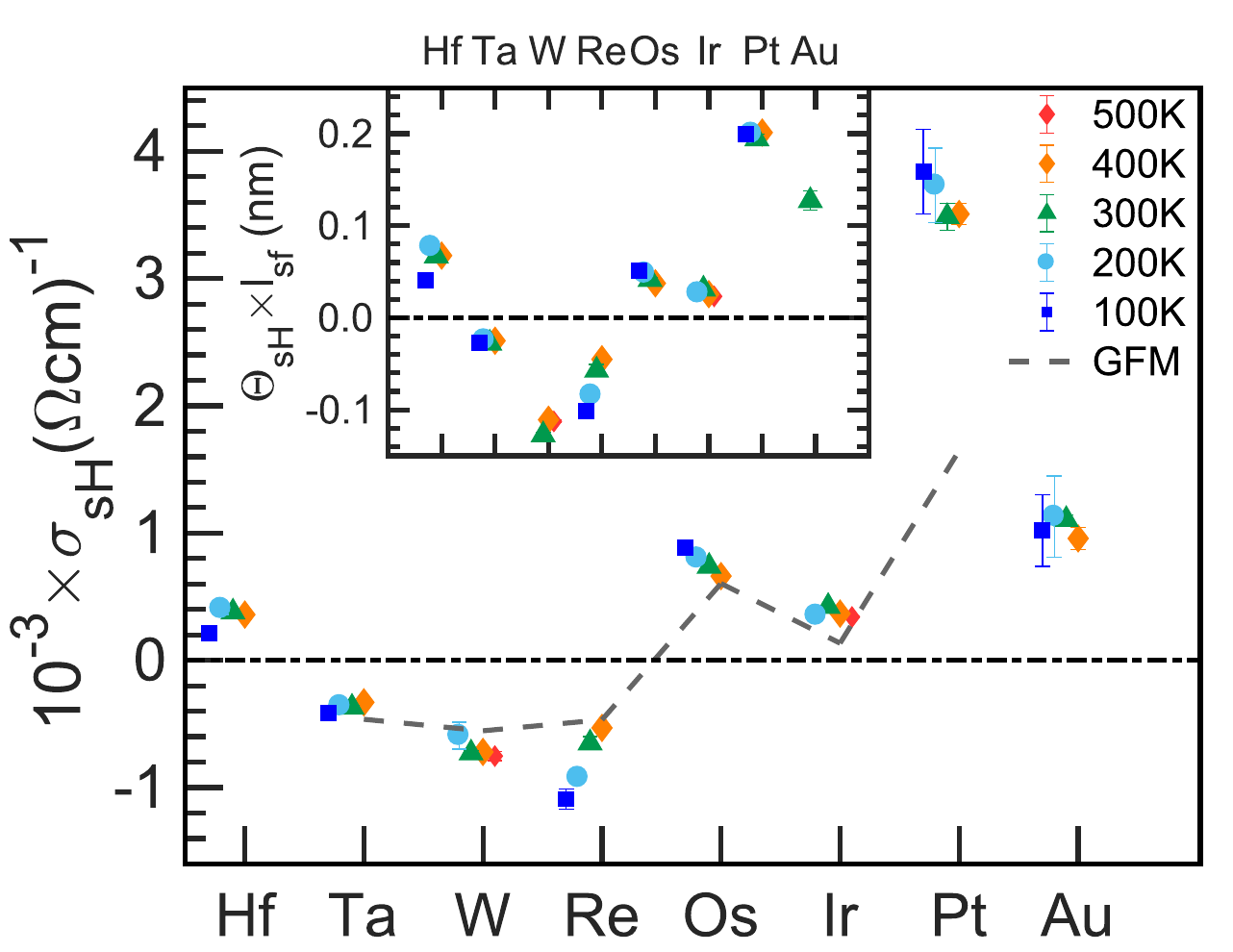} 
\caption{The spin Hall conductivity $\sigma _{\rm sH}$ ($= \Theta_{\rm sH} \times \sigma$) calculated at different temperatures  compared to the room temperature spin Hall conductivity for all 5$d$ elements (except Hf and Au) reported by Tanaka {\em et al.} \cite{Tanaka:prb08} using a Green function method (GFM) and a phenomenological ``quasiparticle damping rate'' to account for disorder. The inset shows the product $\Theta_{\rm sH}\times l_{\rm sf}$ as a function of temperature. Only one value is plotted for Au based on an extrapolation of $l_{\rm sf}$ to RT. For hcp metals, the $c$-axis values are shown. 
}
\label{Fig5}
\end{figure}

The 5$d$ metal spin Hall conductivity (SHC) $\sigma _{\rm sH} = \Theta_{\rm sH} \times \sigma$ is shown for four different temperatures in \cref{Fig5} and is seen to be weakly dependent on temperature, except for Re. The element ($Z$) dependence is in qualitative agreement with the linear response calculations by Tanaka {\it et al.} who used an empirical ``quasiparticle damping parameter'' to represent disorder \cite{Tanaka:prb08}. In these latter calculations, the correct bcc, fcc and hcp structures were used for each element but Hf and Au were not considered. Based upon a rigid band study for the fcc structure, Guo {\em et al.} identified a switch from positive values of $\sigma _{\rm sH}$ for band-filling corresponding to Pt to negative values for band-fillings corresponding to Ta and W \cite{Guo:prl08} and this has been interpreted in terms of Hund's third rule \cite{Morota:prb11}. 
We reproduced the peaks observed by Guo {\it et al.} as a function of band filling for the fcc structure in \cite{Wesselink:prb19} and note that they originate in regions of the Brillouin zone with degeneracies that are strongly affected by SOC. The energies and numbers of these degeneracies depend on the lattice structure and explicit calculation for the bcc and hcp structures shows that the correspondence between Guo's rigid band picture for the fcc structure and observations for materials with other structures is accidental \cite{Nair:tbp21}. For example, it predicts a small, negative SHC for Hf. However, for hcp Hf we find a small, positive value of $\sigma_{\rm sH}$ in agreement with recent experiments \cite{Fritz:prb18, footnote4}.

Our finding of a weak temperature dependence of $\sigma_{\rm sH}$ is in qualitative agreeement with the temperature independence found for Pt by Isasa {\em et al.} \cite{Isasa:prb15a, *Isasa:prb15b} who, however, reported a substantially lower value of $ \sigma_{\rm sH} \sim 1200 \, (\Omega {\rm cm})^{-1}$ (note that we use the $\hbar/2e$ convention \cite{Stamm:prl17}). A subsequent measurement by the same group yielded a room temperature spin Hall conductivity of $\sim 3200 \, (\Omega {\rm cm})^{-1}$ \cite{Sagasta:prb16} in excellent agreement with our value. At low temperatures in clean samples, they observed an enhancement of $\sigma_{\rm sH}$ with decreasing temperature. Such an enhancement was recently found \cite{Xiao:prb19} in calculations in which phonon modes of Pt were explicitly populated as a function of temperature \cite{LiuY:prb15} and it was pointed out that the temperature independence of $\sigma_{\rm sH}$ that we find is characteristic of the classical equipartition approximation \cite{Xiao:prb19}. When the resistivity is no longer linear in temperature, our description of thermal disorder in terms of a Gaussian distribution of uncorrelated atomic displacements is not suitable for studying the weak scattering limit. This is typically well below $100\,$K \cite{HCP90}.

In view of the proportionality of $\Theta_{\rm sH}^{\rm Pt}$ to the resistivity $\rho$ \cite{WangL:prl16} and of $l_{\rm sf}^{\rm Pt}$ to the conductivity \cite{LiuY:prl14, LiuY:prb15}, their product is expected to be independent of temperature. $\Theta_{\rm sH}\times l_{\rm sf}$ is shown in the inset to \cref{Fig5} for all 5$d$ elements for a number of temperatures between 100 and 500 K. As a function of $Z$, it is seen to follow the trend of $\sigma_{\rm sH}$ and as a function of temperature it is indeed approximately constant for all metals except Rhenium for which additional theoretical and experimental studies are desirable.
It remains to be seen how $\Theta_{\rm sH}\times l_{\rm sf}$ behaves when different scattering mechanisms are present simultaneously, in particular interface scattering and bulk thermal disorder. 

{\color{red}\it Conclusions.---}
We have presented a comprehensive ab-initio study of two important spin-orbit coupling related transport parameters in the 5$d$ metals as a function of temperature verifying the generality of the Elliot-Yafet mechanism and establishing numerical benchmarks for experiment. The values of $\rho \, l_{\rm sf}$ , $\sigma_{\rm sH}$ and  $\Theta_{\rm sH} \, l_{\rm sf}$ calculated in this work can be directly used to predict the values of $l_{\rm sf}$ and $\Theta_{\rm sH}$ for the most experimentally relevant  temperatures. In particular, the direct correspondence between the spin-flip diffusion length and density of states at the Fermi level implies that spin-flip scattering can be effectively controlled by alloying whereas a high spin Hall conductivity may be achieved by tuning the Fermi level with respect to degeneracies at high symmetry points. 
Our results indicate that the magnitude of the SOC is not the sole determinant of $l_{\rm sf}$ and $\Theta_{\rm sH}$ but that crystal structure and associated details of the electronic structure are just as important. 

{\color{red}\it {Acknowledgements.---}}This work was financially supported by the ``Nederlandse Organisatie voor Wetenschappelijk Onderzoek'' (NWO) through the research programme of the former ``Stichting voor Fundamenteel Onderzoek der Materie,'' (NWO-I, formerly FOM) and through the use of supercomputer facilities of NWO ``Exacte Wetenschappen'' (Physical Sciences). R.S.N (project number 15CSER12) and K.G. (project number 13CSER059) acknowledge funding from the Shell-NWO/FOM ``Computational Sciences for Energy Research (CSER)'' PhD program. The work was also supported by the Royal Netherlands Academy of Arts and Sciences (KNAW). Work in Beijing was supported by the National Natural Science Foundation of China (Grant No. 61774018), the Recruitment Program of Global Youth Experts, and the Fundamental Research Funds for the Central Universities (Grant No. 2018EYT03).


\begin{thebibliography}{63}
\bibitem {footnote1}%
  \BibitemOpen
  \href@noop \bibinfo {{\color{blue}We could calculate the
  conductivity entirely from first principles; see \cite{Savrasov:prb96b,
  LiuY:prb15} but because the results are not always in as good agreement with
  experiment as in \cite{LiuY:prb15} (and the ordering temperature of magnetic
  materials is underestimated by spin-wave theory), it is more expedient to
  match the thermal disorder to reproduce the experimental resistivity (and
  magnetization) as in \cite{LiuY:prb15, WangL:prl16, Starikov:prb18,
  Wesselink:prb19, Gupta:prl20}.}}\BibitemShut {Stop}%
\bibitem {footnote4}%
  \BibitemOpen
  \href@noop \bibinfo {{But at variance with
  explicit calculations for hcp Hf \cite{Freimuth:prl10, Fritz:prb18}. When we
  calculate the spin Hall conductivity as a function of band filling using two
  different methods \cite{WangL:prl16, Wesselink:prb19} and \cite{Qiao:prb18},
  we find that the SHA changes sign very close to the Hf bandfilling making it
  very sensitive to details of the Hf band structure.}}\BibitemShut {Stop}%
\end{thebibliography}

%

\end{document}